\documentclass [12pt, draftclsnofoot, onecolumn]{IEEEtran}
\usepackage{graphicx}
\usepackage{caption}
\usepackage{subcaption}
\usepackage{stmaryrd}
\usepackage{verbatim}
\usepackage{amssymb}
\usepackage{amsmath}
\usepackage{amsfonts}
\usepackage{mathrsfs}
\usepackage{pst-node}
\usepackage{mathabx}
\usepackage{wasysym}
\usepackage{amsthm}
\usepackage{cite}
\usepackage{fixltx2e}
\usepackage{float}
\usepackage{dblfloatfix}
\usepackage[ampersand]{easylist}
\usepackage[ruled,vlined]{algorithm2e}
\usepackage{afterpage}
\usepackage{multirow}
\usepackage{tabu}
\DeclareMathAlphabet{\mathpzc}{OT1}{pzc}{m}{it}

\begin{document}
\title{\LARGE{Beamforming Learning for mmWave Communication: Theory and Experimental Validation}}
\author{ Mohaned Chraiti,~Dmitry Chizhik,~Jinfeng Du,~Reinaldo A. Valenzuela,~Ali Ghrayeb and Chadi Assi}
\maketitle
{\let\thefootnote\relax\footnotetext{M. Chraiti and C. Assi and are with Concordia University, Montreal,
Canada (emails:m\_chrait@encs.concordia.ca and assi@ciise.concordia.ca).
\par D. Chizhik, J. Du  and R.A. Valenzuela are with Nokia Bell Laboratories, NJ, USA, (e-mails:dmitry.chizhik@nokia-bell-labs.com, jinfeng.du@nokia-bell-labs.com and reinaldo.valenzuela@nokiabell-labs.com).
\par A. Ghrayeb is with the ECE Department, Texas A$\&$M University at Qatar, Doha, Qatar (e-mail: ali.ghrayeb@qatar.tamu.edu).
\par This work was supported by an NSERC Discovery Grant, FQRNT Grant and Concordia University. The statements
made herein are solely the responsibility of the authors.}}
\begin{abstract}
 To establish reliable and long-range millimeter-wave (mmWave) communication, the high propagation losses of this frequency range must be overcome. Beamforming is deemed to be a promising solution. Although beamforming can be done in the digital and analog domains, both approaches are hindered by several constraints when it comes to mmWave transmission and reception. For example, performing fully digital beamforming in mmWave systems (particularly on mobile devices) involves using many radio frequency (RF) chains, which are bulky, expensive and consume high power. This necessitates finding more efficient ways for using fewer RF chains while taking advantage of the available large antenna arrays. One way to overcome this challenge is to employ (partially or fully) analog beamforming through proper configuration of phase-shifters. Existing works on mmWave analog beam design either rely on the knowledge of the channel state information (CSI) per antenna within the array, require a large search time (e.g., exhaustive search techniques) or do not guarantee a minimum beamforming gain (e.g., codebook based beamforming techniques). In this paper, we propose a beam design technique that reduces the search time and does not require CSI while guaranteeing a minimum beamforming gain. The key idea of the proposed scheme derives from observations drawn from measurements obtained through real-life measurements. It was observed that for a given propagation environment (e.g., inside the coverage area of a mmWave base station) the azimuthal angles of dominant signals could be more probable from certain angles than others. Thus, beam directions for future connected users can be predicted from the measurements collected by previous connected users. In fact, such measurements are used to build a beamforming codebook that regroups (i.e, clusters) the most probable beam designs containing dominant signals. We invoke Bayesian machine learning for measurements clustering. We evaluate the efficacy of the proposed scheme in terms of building the codebook and assessing its performance through real-life measurements. We demonstrate that the training time required by the proposed scheme in only $5\%$ of that of exhaustive search. This crucial gain is obtained while achieving a minimum targeted beamforming gain.

\end{abstract}
\section{Introduction}
\subsection{Motivation}
Due to the ever increasing market demands for ultra high rate wireless links with ubiquitous connectivity, the wireless industry is moving towards using millimeter wave (mmWave) frequencies, that offer large bandwidth, on the order of GHz. However, mmWave communication is limited by the physical properties of the channel, which has been shown to be sensitive to blockage (e.g., human body could cause up to 40 dB of power loss) and to have high path loss. To compensate the high propagation losses,  mmWave transmitter/receiver are anticipated to be equipped with large number of antennas that would enhance transmit/receive antenna gain. Integrating a large antenna array into small wireless devices, such as mobile phones, is also feasible due to the small size of mmWave antennas \cite{antennammWave}.

\par Large antenna arrays have the potential to provide considerable gains in the received power by using beamforming techniques. Providing the appropriate beam design, however, is hindered by several challenges. Due to the high-power consumption and cost of mmWave radio-frequency (RF) chains, it is anticipated that mmWave mobile devices will be equipped with a large number of antennas but fewer RF chains \cite{mmWave0,mmWave1,mmWave2,mmWave3,mmWave4,mmWave10,mmWave11,mmWave12,mmWave13}. Consequently, performing fully digital baseband beamforming may not be possible to realize in a mobile device. 

\par Several works have been published on the subject of mmWave beamforming where a large antenna array and fewer RF chains are considered \cite{mmWave0,mmWave1,mmWave2,mmWave3,mmWave4,mmWave10,mmWave11,mmWave12,mmWave13}. The authors rely on analog beamforming where beams are made through the configuration of low-cost phase-shifters. While some of them suggested using analog beamforming exclusively, others considered analog-digital hybrid beamforming. However, no matter which operation mode is considered, analog beamforming is an integral part of future mmWave devices and developing efficient techniques to configure the antenna phase-shifters is required. In this context, two main approaches have been proposed, namely precoding and beam training.

\par For precoding \cite{mmWave0,mmWave1,mmWave2,mmWave3,mmWave4}, the authors rely on the knowledge of the channel state information (CSI) associated to \textbf{each antenna within the array}, to compute the phase-shifters. Several solutions have been proposed where different objectives were considered. In \cite{mmWave0}, a precoding algorithm was developed to minimize the mean-squared error at the receiver, while in \cite{mmWave1,mmWave2,mmWave3}, the authors focused on the beam designs that enhance the achievable rate. To reduce the computational complexity, and to lower the energy consumption, the authors exploited the sparsity of the mmWave channel matrix. They formulated the problem as a sparse approximation problem. Then, they used sub-optimal low-complexity techniques, such as compressive sensing, to solve the problem. They showed that the proposed techniques achieve near optimal performance in terms of beamforming gain. For more energy efficiency, the authors in \cite{mmWave4} considered a sub-connected architecture, i.e, not each antenna is connected to each RF chain. They showed that such architecture increases the energy-efficiency while achieving almost similar performance as that of a fully-connected architecture as considered in \cite{mmWave1,mmWave2,mmWave3}.

\par Although the earlier cited techniques achieve near maximum array gain, they rely heavily on the full knowledge of the CSI associated with each of the antennas. Acquiring such knowledge could require large overhead and is time consuming, especially when the number of antennas is large as expected in next generation cellular systems. Moreover, while the receiver may use pilot symbols to estimate CSI and to perform receive beamforming, transmit beamforming may require feeding back the CSI from the receivers, which induces a considerable overhead as compared to acquiring CSI at the receiver.\footnote{In frequency division based systems (e.g., LTE), devices transmit and receive signals over different frequencies and hence the CSI of the uplink channel differs from the one in the downlink channel. Performing beamforming at the transmitter requires CSI that is estimated by the receiver and fed back to the transmitter.} This renders CSI-based beamforming approaches practically undesired for transmit beamforming \cite{mmWave10,mmWave11,mmWave12,mmWave13}. In addition, in mmWave communication, the received signal per antenna before beamforming is expected to be very weak. As the channel coefficients have to be estimated per antenna, large errors in channel estimation will likely to be encountered and this will lead to limiting the gains achieved through beamforming \cite{mmWave11}.

\par Designing analog beams without the knowledge of the CSI per antenna is the main motivation behind the development of the second approach, namely, beam training \cite{mmWave10,mmWave11,mmWave12,mmWave13}. The idea is to steer a beam in different directions, according to a predetermined beamforming codebook, then choose the one that maximizes the received signal power.\footnote{Note that using beam training techniques, a receiver will get a scalar product of the channel coefficients and the phase-shifters weight, all multiplied by the transmitted symbol. Therefore, although beam training techniques do not require estimating the CSI for each antenna while designing the analog beam, the receiver may need to estimate the aforementioned scalar product, which is the equivalent of estimating one channel coefficient, to be able to decode the transmitted symbol. This is similar to the case when the receiver is equipped with one antenna. We stress here that such information is not needed while performing transmit beam training.} A naive beamforming training technique is done through exhaustive search by considering a narrow beam, rotating the beam in small steps and then choosing the one that maximizes the received power. Exhaustive search could achieve the highest array gain, if the used codebook is of high resolution (a narrow beam and a small rotation beam step). However, in this case, beam training becomes time consuming. In an effort to reduce the search time, hierarchical beam training has been proposed \cite{mmWave10,mmWave11,mmWave12,mmWave13}. The authors suggested to use a divide-and-conquer search process across the codebook levels where at each level, the best beam contained in the higher-level beam (i.e., lower resolution level) with the largest gain is selected.
\par The main drawback of hierarchical beam training is the absence of minimum gain guarantees such as achieving a gain within a certain gap to the maximum. In fact, at a low resolution level of the codebook, a particular wide beam could have the highest gain, however, there is no guarantee that one of its descendant beams will achieve the highest gain or at least a gain within a certain range. Moreover, the choice of key parameters of hierarchical beam training (e.g., number of levels, widths of the beams in each level, etc.) is not justified, meanwhile they heavily impact the beamforming gain and the beam search time. Furthermore, contrary to exhaustive search where the received power from the directions of the dominant signals is much higher than the remaining ones, two beams with low resolution level could have somewhat similar gains while considering hierarchical search. This makes the decision about the beam with largest gain subject to error when measurements lack accuracy. Meanwhile, increasing the measurements accuracy requires longer time that could be as high as the exhaustive search time. Motivated by this, we aim in this paper to provide an analog beamforming technique that considerably reduces the search time and does not require the CSI while guaranteeing a minimum beamforming gain.

\begin{figure*}[h]
    \centering
    \begin{subfigure}[b]{0.4\textwidth}
        \includegraphics[scale=0.5]{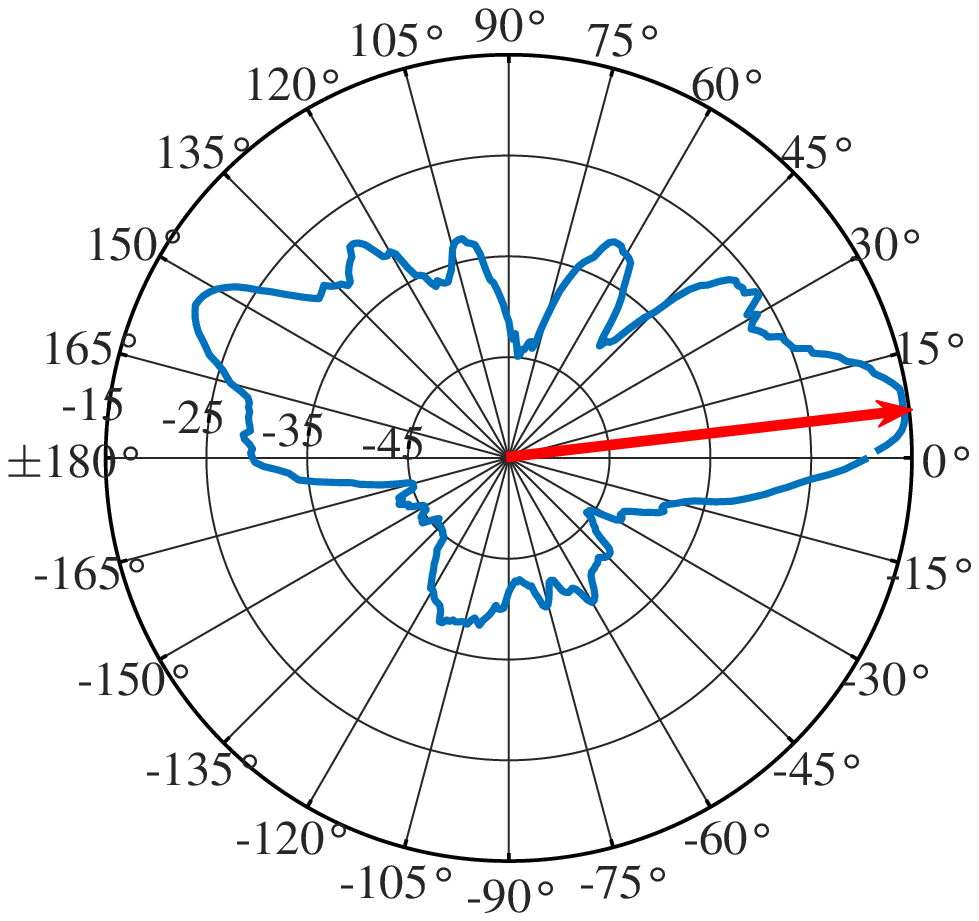}
        \caption{Office 1.}
        \label{Office1}
    \end{subfigure}
\begin{subfigure}[b]{0.4\textwidth}
        \includegraphics[scale=0.5]{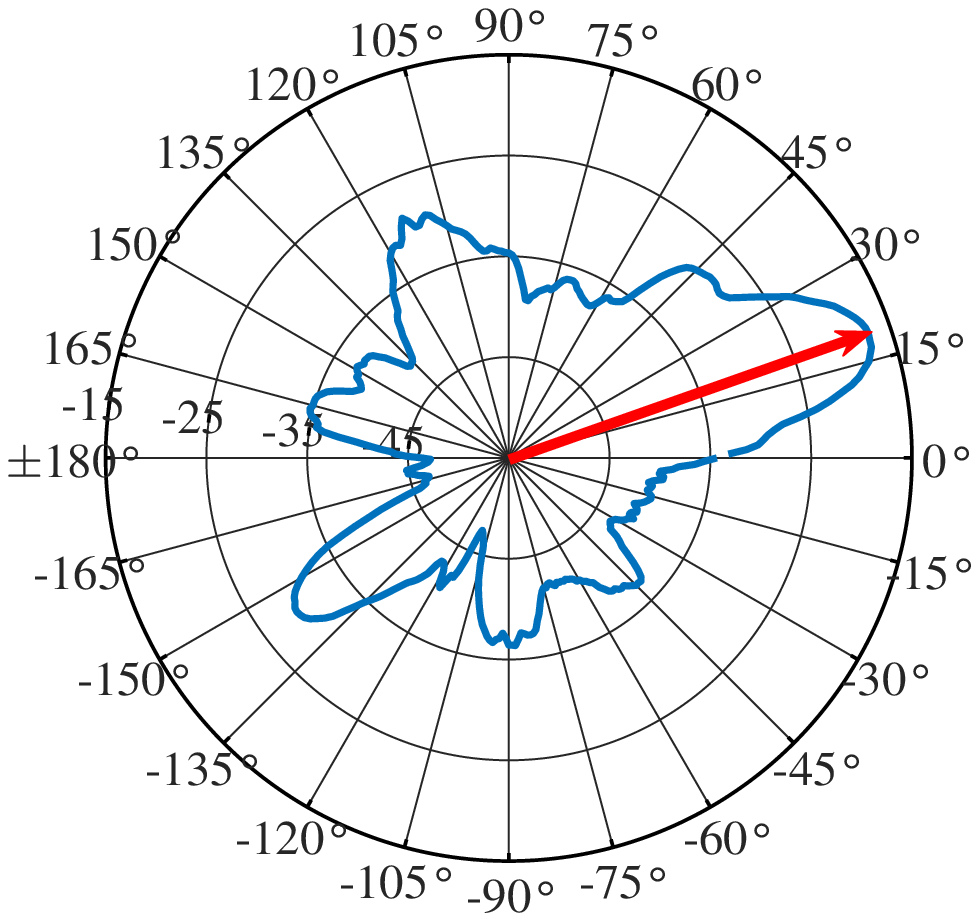}
        \caption{Office 2.}
        \label{Office2}
    \end{subfigure}
\label{S1Dirichlet}
    \caption{Samples of azimuthal radiation patterns inside offices.}\label{S1planing}
\rule{\textwidth}{0.5pt}
\end{figure*}

\subsection{Proposed Solution Overview}

\par The idea of the proposed approach derives from a key observation of real world mmWave measurements that we conducted at Bell Labs, Crawford Hill, NJ. We observed that, in a given propagation environment (e.g., inside the coverage area of a mmWave base station), the azimuthal Angle-of-Arrival(AoA)/Angle-of-Departure (AoD) of dominant signals could be more probable from certain angles than others, i.e., there is a possible similarity among azimuthal radiation patterns associated with dominant signals. One of the conducted experiments was in an indoor office environment and it consisted of placing a 28 GHz transmitter in the corridor and a horn antenna receiver in different offices and at different positions in the same office. The receiver was mounted on a rotating platform and was able to measure signals received from all azimuthal angles with one degree precision. The receiver was placed randomly in each of the offices. The experiment was also conducted in the presence of humans and other obstacles such as experimental equipment and furniture (more details on the used equipment and experiment parameters are provided in Sec. \ref{Results}.) A sample of azimuthal radiation patterns obtained at two different offices are depicted in Figs. \ref{Office1} and \ref{Office2} where we can observe some similarity between the two radiation patterns. In Figs. \ref{Office1} and \ref{Office2}, the dominant signals' directions are somewhat related to the physical direction of the office doors. This similarity is expected, given that mmWave signals have poor penetration and hence dominant signals' AoA/AoD are somewhat related to the physical architecture of the propagation environment. It also aligns with the mmWave channel sparsity that suggests that only few dominant multipath components (that are strongly related to the physical architecture of the environment) contribute to the received signal power, and hence there is a higher probability of having similarity among them as compared to rich scattering experienced at lower frequencies.\footnote{The hypothesis claiming that the angles of dominant mmWave signals take random values (i.e., uniformly distributed in $[0^{\circ}, 360^{\circ}]$) and hence are independent from the physical architecture contradicts with the existing works on mmWave such as the ones that suggest using the angle of received signal to estimate the locations of mmWave devices (i.e., angle of propagation paths are correlated to certain locations) \cite{LocalIndoorTechn,LocalIndoorTechn1,LocalIndoorTechn2}.}


\par  The possible similarity between the AoA/AoD of dominant signals suggests that there would be a possible similarity between the beams design of previously and future connected users in the same propagation environment, e.g., located in the same coverage area of a mmWave base station (BS). The proposed beamforming technique is hence based on exploiting the experience (i.e., beams design) of previously connected users to predict beam designs for future connected users. The proposed approach consists of collecting measurements a priori from users or by the service provider to build a codebook that regroups the most probable analog beams containing dominant signals. The codebook is built while taking into consideration a constraint on the minimum beamforming gain. Once the codebook is set up, the transmitter/receiver steer the beam according to the codebook, then choose the one that maximizes the received signal power.

\par The beam search time is mainly determined by the codebook size, i.e., a smaller codebook gives a shorter search time. Therefore, the objective of this work is to minimize the size of the beamforming codebook subject to a minimum guaranteed gain. Building such codebook from measurements, however, could be challenging since the collected measurements are discrete and of large size. Moreover, there are multiple parameters to determine such as the codebook size and beams' directions. These are in addition to the constraint on the minimum guaranteed performance. The problem at hand gives rise to a mix of discrete-continuous optimization problem with a large search space, which is unclear how to solve through optimization techniques, and it may not even be scalable.

\par As the key idea is to exploit similarity among beams, the problem at hand can be seen as a clustering problem where approximately similar beams are grouped together and one beam is delegated to represent each of these clusters. The delegated beams will be the elements of the codebook. The fact that we have a clustering problem and a huge size of data to process makes machine learning a potential candidate to infer the codebook \cite{MachineBook1,DirichletBook1,DirichletBook2,DirichletBook3}. However, there are multiple challenges that need to be taken into consideration. First, the optimal codebook is unknown, which suggests that the technique should be unsupervised. Second, the size of the codebook is also unknown and hence the used method should be nonparametric.\footnote{Nonparametric machine learning techniques are those that do not require the number of clusters as input. For instance, machine learning techniques based on K-mean require setting a priori the number of clusters $K$ and hence they can not be considered as nonparametric clustering techniques \cite{MachineBook1,datamining}.} Third, the proposed technique should offer the ability to auto-update the codebook if more measurements are available or when the physical environment changes due, for instance, to constructions. All those criteria are met by the well known Bayesian machine learning approach, and hence it will be considered in this paper to solve the problem at hand \cite{MachineBook1,DirichletBook1,DirichletBook2,DirichletBook3,Bouguila1,Bouguila2,Bouguila3,Bouguila4}. 


\par The core idea of the machine learning approach adopted in this paper is that the codebook parameters (beams' widths and directions) are treated as random variables, which naturally correspond to some joint probability distribution conditioned on the measurement points (i.e., observations) \cite{MachineBook1,DirichletBook1,DirichletBook2,DirichletBook3,Bouguila1,Bouguila2,Bouguila3,Bouguila4}. The parameters that maximize this conditional probability distribution are learned (i.e., inferred) from the observations. The inference process may be summarized as follows. We define the probabilistic model that binds the measurements to the codebook parameters while considering the constraints at hand. This has to be done in such a way that we can infer the codebook parameters from the parameters of the probabilistic model of the measurements in hand. We make use of Gibbs sampling theory and Bayes' theory to infer the conditional probability (called, the posterior) and the parameters that maximize it, and this is used to obtain the codebook parameters \cite{BouguilaGibbs}.

\subsection{Contributions}

\par In this paper, we make multiple contributions. We first propose a novel system design for beamforming prediction and describe the communication process among the system elements, namely, the users, the service provided and the BS. The process contains three major steps, namely, measurements collection, codebook building and beam training. Second, we provide a measurement-based codebook design technique using Bayesian machine learning where the problem is formulated and solved. The proposed codebook guarantees a minimum gain. It is worth mentioning that, to the best of our knowledge, we are the first to exploit measurements of previously connected users to predict the beam design for future connected users.\footnote{In low frequencies used in 4G systems and lower generations, the obstacle penetration depth is much higher than the one of mmWave frequencies. This makes the AoA/AoD of dominant signals in different devices much less correlated to each other, if not completely independent. Nonetheless, the widely considered model used to characterise the channel effect such as Rayleigh and Nakagami are a clear proof of the independence between the users' AoA/AoD \cite{PlanningBook2}.} Third, we conducted real-life experiments, considering indoor office environments, to validate the proposed approach and show its efficacy. We show that the proposed approach achieves the intended goal while saving more than $95\%$ of the search time as compared to exhaustive search. Moreover, we show that the proposed approach could exploit multiple side information such as the positions of users to reduce the search time. Indeed, it uses the available side information to clusters the users in smaller sets with higher azimuthal beams similarity.

\par The rest of the paper is organized as follows. In Section \ref{Preliminary}, preliminaries on nonparametric Bayesian statistics are provided. In Sections \ref{Sysmodel} and \ref{Inference}, the system design and the codebook inference process are provided, respectively. The inference algorithm is described in detail in Section \ref{ModelInf}. We discuss the proposed technique process in the case where multiple side information are available \ref{extension}. The performance of the proposed approach is assessed and compared to existing benchmark approaches in Section \ref{Results}. We conclude the paper in Section \ref{Conclusion}.

\section{Background on Bayesian Statistical Learning}\label{Preliminary}
Bayesian learning is different from other commonly used machine learning techniques such as deep neural networks, random forest, reinforcement learning, etc. \cite{MachineLearning}. In fact, the Bayesian method is statistics based and is known to be analytical in nature, although the solution is obtained algorithmically \cite{DirichletBook1,DirichletBook2,DirichletBook3,BouguilaGibbs}. This stems from the fact that the inference algorithm is used to obtain an approximation of (i.e., learn) the conditional probability of the intended parameters given the observations. Moreover,
it has been proven theoretically that the results converge to the exact intended result (called, true posterior) as the number of observations increases \cite{DirichletBook1,DirichletBook2,DirichletBook3,BouguilaGibbs}. It also offers the possibility to characterize probabilistically the gap to the true posterior.

\par Furthermore, other machine learning approaches may require pre-fixing some parameters (e.g., the number of clusters) and/or consider that the parameters take values in finite discrete parameters space (e.g., reinforcement learning) \cite{MachineLearning}. This makes them not suitable for our case since the number of clusters, which will reflect the size of the codebook, is unknown a priori and may vary from one environment to another. In addition, the elements of the codebook could take values in a space of infinite elements. For instance, the beam direction could take any value in the angular interval $[0^{\circ},360^{\circ}]$. Nonetheless, it is not clear how one can apply any of the machine learning techniques to solve the problem at hand.


\par In Bayesian statistics, any form of uncertainty is expressed as randomness. Therefore, we model the intended unknown parameters (i.e., codebook parameters), denoted by $\Theta=\{\theta_i,i\in\mathbb{N}\}$, as random variables, and they take values in space $\Omega$. The observations $\{x_1,\ldots,x_n\}$ are assumed to be generated in two stages. First, the parameters are sampled from a space $\Omega$ according to a prior distribution $G_0$. The prior gives the possibility to incorporate our thoughts, experiences, knowledge, etc, in how the parameters of the model should look like. For example, in the case where the transmitter and the receiver are both located in the corridor, from our experience, dominant signals most likely come from the direction of the transmitter. Second, the data is independently sampled from the distribution $P_{\Theta}$. That is,
\begin{equation}
\begin{aligned}
&\,\,\,\,\,\,\,\,\,\,\,\,\,\,\,\,\,\,\,\Theta\sim G_0 \\
&x_1,\ldots,x_n|\Theta\sim_{iid}P_{\Theta}.
\end{aligned}
\end{equation}


\par The objective now is to draw a conclusion about the values of $\Theta$ from the observations, which is provided through inferring the posterior distribution $G(\Theta)\overset{\Delta}{=}P[\Theta|x_1,\dots,x_n]$ from which the most likely value of $\Theta$ is extracted. Using Bayes' rule, we have
\begin{equation}
\begin{aligned}
G(\Theta\in\Omega)&=\frac{\prod_{i=1}^np(x_i|\Theta)G_0(\Theta)}{\int_{d\Theta\in\Omega}\prod_{i=1}^np(x_i|d\Theta)G_0(d\Theta)}.
\end{aligned}
\end{equation}
In almost all scenarios, the explicit expression of the posterior is difficult, if not possible, to provide analytically. Nonetheless, to obtain the posterior and the values of the elements of $\Theta$, there are inference approaches that can be used, such as Gibbs-sampling \cite{DirichletBook1,DirichletBook2,BouguilaGibbs} (more on this in Sec. \ref{ModelInf}.)

\par An inference model is said to be parametric if the space of the parameters $\Omega$ has a finite dimension $K$ that is known a priori. Obviously, this model is not suitable for our case since the number of parameters, which is essentially related to the size of the codebook, is unknown. In this case, $\Omega$ has to be of infinite dimensions and thus the inference model is said to be nonparametric, which is what we consider in this paper.


\section{System Design}\label{Sysmodel}
We consider a generic model that consists of a mmWave BS serving multiple users within its coverage. Each of the users' device is assumed to be equipped with multiple antennas and few mmWave RF chains (could be as small as one). Users will perform receive/transmit beamforming to enhance, respectively, their received power and the one received by the BS. This includes their ability to point the beam approximately in any possible azimuth direction and to perform reasonably narrow and wide beams as intended, e.g., in the range of $10^{\text{o}}$ to $60^{\text{o}}$. Although the proposed approach can be used for beam prediction at the users' devices as well as at the BS, we focus in this paper on the users' devices and the same process applies to the BS.

\par This paper presents a novel mmWave analog beamforming technique that does not require CSI and provide a minimum gain guaranteed. The key idea is to exploit the possible similarity among dominant signals' AoA/AoD of different users in the same propagation environment ( see Figs. \ref{Office1} and \ref{Office2}.) Especially, we make use of measurements collected a priori from previously connected users and/or by the service provider to build a codebook regrouping the most probable beams that contain dominant signals. Future connected users hence will consult the already-built codebook then pick the beam that maximizes the received power. The aforementioned phases, namely, collecting measurements, building the codebook and beam training, are briefly described in the following.

\subsection{Collecting Measurements}
Measurements are collected from different locations in the coverage area of the BS. At each position, the task consists of establishing a narrow beam then rotating it in small steps. For each position, the received signal power and its associated direction is recorded and shared with the BS. The task of collecting measurements to build the codebook for receive beamforming differs slightly from the one for transmit beamforming. As for the first one, we suggest that the measurement will be collected by a mobile device then sent to the BS, whereas for transmit beamforming, the mobile device performs transmit beamforming at different directions and the BS records the received signals. In the rest of the paper, measurements correspond to the set of angles (AoA/AoD) and their associated received power.

\par The measurements could be collected by the service provider as well as by the users' devices. We also suggest that the users continue to collect measurements, even after building the beamforming codebook, which could be done, for instance, when the network is not busy (i.e, low traffic) and when they are idle. This will help to enhance the codebook accuracy, since it is intuitive that the more observations we have, better inference accuracy will be obtained (see Sec. \ref{Inference} for more details). This also gives the advantage of updating the codebook when changes in the environment occur (e.g., constructions, tree leafs loss), without the intervention of the service provider.

\subsection{Codebook Building}
 Given the measurements, the BS builds a beamforming codebook considering a minimum performance criterion, that is, achieving a gain within a certain gap to the maximum for almost all cases. In other words, using the codebook, the probability of having a gain within a gap (denoted by  $\gamma$) from the maximum (denoted by $\text{Max}_{\text{Gain}}$) is desired to be higher than a certain threshold $O_{th}$ (e.g., $90\%$).\footnote{This constraint has a similar form to commonly used performance criterion such as the outage probability.} This constraint can be formulated analytically as follows.
\begin{equation}
Pr(\text{Gain}\geq \text{Max}_{\text{Gain}}-\gamma)\geq O_{th}.
\end{equation}
The maximum gain that is obtained through an exhaustive search.

\par The proposed approach exploits the similarity among the beams containing intended dominant signals, i.e., beams with a gain above the threshold $\text{Max}_{\text{Gain}}-\gamma$. To do so, we make use of Bayesian learning to cluster theses beams. We then extract the elements of the codebook from the obtained clusters. Each element (i.e., training beam) will be defined through two parameters, namely, direction and width. It is to note that the channel responses may change from a coherence bandwidth to another. Therefore, we propose to build a codebook per coherence bandwidth.

\subsection{Beam Training}\label{Errordirec}
When a user attempts to establish a communication with a BS, the latter shares with the user the beam training codebook. The user steers receive/transmit beamforming according to the element of the codebook, then picks the beam design that maximizes the received signal power. Here, beam training for transmit and receive beamforming differs slightly from each other. In fact, the beam selection is made by the user for receive beamforming, whereas it is made by the BS for transmit beamforming where the BS feeds back the index of the best beam design.

\par Using the proposed training technique, the user has to orient the beams as indicated by the codebook. Here, there is an underlying assumption concerning a reference direction (i.e., direction $0^{\text{o}}$) that should be known by the BS as well as by the users and has to be the same one used to build the codebook. To elaborate, let us consider the case where one of the codebook elements indicates that the beam direction is $90^{\text{o}}$. This suggests that the user has to know first the reference direction $0^{\text{o}}$, then orients the beam $90^{\circ}$. There are multiple ways to set a reference direction \cite{LocalBookCompass,LocalBookGPS,LocalBookGPSPrecision,localBook1indoor, localBook1indoor1,LocalIndoorTechn, LocalIndoorTechn1,LocalIndoorTechn2,LocalIndoorTechn3}. For instance, it could be one of the geodetic directions such as the true north which can be easily obtained through the digital compass of the user device\cite{LocalBookCompass}. Another option is to consider the user-BS direction as a reference direction. In this case, we suggest that the BS share its position with the user, who will in turn use a Global Positioning System (GPS) to identify its position and then the user-BS direction \cite{LocalBookGPS,LocalBookGPSPrecision}. Although this solution is more applicable for outdoor communications, some highly precise solutions and products have been proposed and commercialized including the interior positioning system (IPS) \cite{LocalNokiaIPS, LocalNokiaIPS1}.


\par Random errors along with the estimation of the reference direction are expected to occur. In practice, errors induced by the above listed solutions are reasonably low. In fact, nowadays, a phone compass has a margin of incertitude of $5^{\text{o}}$ for almost all case scenarios. Moreover, a study made by the government of the United-States showed that the margin of incertitude of the GPS is less than $8$ meters for $95\%$ of the cases \cite{LocalBookGPSPrecision}. An example of the effect of the GPS precision on the direction estimation error is depicted in the following. Let us assume that there is a user located $50$m away from the BS. In this case, the error in estimating the user-BS direction is less than $9^{\text{o}}$ with probability higher than $95\%$. The error becomes less than $5^{\text{o}}$ when the user is $100$m away from the BS. Nonetheless, such error could be handled by the proposed approach, as it will be shown in Section \ref{Results}. For instance, along with the measurements collection process, there is a random error in the reference direction that can reach up to $20^{\text{o}}$. Nonetheless, the provided results show that the proposed approach is robust against the error in the reference direction estimation. 

\section{Codebook Inference}\label{Inference}
The primary objective of our work is to infer a beamforming codebook that regroups the most probable beams directions and widths that would meet a given performance criterion. For clustering, we make use of the nonparametric Bayesian approach, since the size of the codebook is unknown and the used technique has to be unsupervised. This method consists of inferring a probabilistic model on the measurements. Here, the codebook elements have to correspond to or computed from the parameters of the inferred model. Therefore, the inference model has to be carefully chosen. The problem formulation as well as the inference method are described in the following.

\subsection{Features Selection}\label{feature}
The codebook will be derived from the measurements through clustering beams that meet the minimum performance criteria, i.e., $\text{Gain}\geq \text{Max}_{\text{Gain}}-\gamma$. Therefore, the first step is to extract from the measurements the beams with the intended gain. Particularly, the BS considers the beams in which each sub-beam has a gain higher than the minimum required. This increases the probability of having a gain higher than the minimum value even if a part of the beam is selected or a beam with a slightly larger width is considered.

\par Each of the considered beams will be defined through two features.\footnote{In machine learning, feature selection consists of selecting the relevant variables from the data before clustering.} The first one consists of the width of the beam whereas the second one consists of the direction of the central ray of the beam. The observations to consider for inference are hence a set of $N$ points each defined through two elements denoted by $\{x_i,y_i\}$ that correspond, respectively, to direction and width.

\subsection{Inference Model: Dirichlet Process}
In this paper, we use a Bayesian approach for inference \cite{BouguilaGibbs,DirichletBook1}. This requires defining a process from which the observations (i.e., measurements) $\{x,y\}$ are sampled. In Bayesian statistical learning, the observations are assumed to be generated through a process that consists of two stages: first, the parameters of the distribution (denoted by $\Theta$) are sampled from certain distributions $G_0(\Theta\in \Omega)$, then the observations are sampled from an obtained distribution (denoted by $P_{\Theta}(x,y)$) that is defined through the sampled parameters. Now, we need to describe in detail the process from which the observations are sampled.

\par Defining a process includes defining the form of the distribution on the measurements. Recall that, through measurements, we showed that there are some beams that are more probable than others (we refer readers to Figs. \ref{Office1} and \ref{Office2}.) Examining the histogram of $\{(x_i,y_i), i\in[1,N]\}$ may reveal peaks with different heights and widths, resembling a mixture of a bivariate (2D) Gaussian distribution, which can be used as an approximate shape of $P_{\Theta}(x,y)$. Particularly, the distribution of the direction of the beams ($x$) is defined through a wrapped Gaussian distribution, given that the angle $x$ is a circular variable \cite{GuaussianBook}.

\par The mixture of distributions is defined through two sets of parameters: the mixture elements' parameters $\boldsymbol{\theta}=\{\theta_1,\theta_2,\ldots\}$ and the mixtures' weights $\boldsymbol{\pi}=\{\pi_1,\pi_2,\ldots\}$, i.e., $\Theta=\{\boldsymbol{\pi},\boldsymbol{\theta}\}$. A bivariate Gaussian mixture has the following form.\footnote{
We note that the parameters to infer in (\ref{eq6}) can be analytically associated to (i.e., computed from) those intended in the codebook process as will be shown in Sec. \ref{parameters}.}

\begin{equation}\label{eq6}
P_{\Theta}(x,y)=\sum_{k}\pi_k\mathcal{N}_{\theta_k}(x,y),
\end{equation}
where $\mathcal{N}_{\theta_k}(x,y)$ is the probability density function (PDF) of the wrapped bivariate Gaussian distribution given the set of parameters $\theta_k$. That is,
\small
\begin{equation}\label{eq7}
\begin{aligned}
\mathcal{N}_{\theta_k}(x,y)&=\frac{1}{\sqrt{2\text{pi}}\sigma_{k,y}}\exp\left(-\frac{1}{2}\frac{(y_{C_k}-y)^2}{\sigma^2_{k,y}}\right)\times\sum_{i=-\infty}^{+\infty} \frac{1}{\sqrt{2\text{pi}}\sigma_{k,x}}\exp\left(-\frac{1}{2}\frac{(x_{C_k}+i\times360-x)^2}{\sigma^2_{k,x}}\right)\\
&=\frac{1}{2\text{pi}\sigma_{k,y}\sigma_{k,y}}\sum_{i=-\infty}^{+\infty}\exp\left(-\frac{1}{2}\left[\frac{(x_{C_k}+i\times360-x)^2}{\sigma^2_{k,x}}+\frac{(y_{C_k}-y)^2}{\sigma^2_{k,y}}\right]\right),
\end{aligned}
\end{equation}
\normalsize
where $\text{pi}\approx 3.14$. Moreover, $(x_{C_k},y_{C_k})$ and $\text{COV}_{C_k}=\begin{bmatrix}\sigma_{k,x}^2 & 0 \\ 0 & \sigma_{k,y}^2\end{bmatrix}$ denote respectively the mean and the covariance matrix of the unwrapped version of the bivariate distribution, i,e., the parameters of the $k$th cluster $\theta_k$ \cite{GuaussianBook}.

\par Since we are considering a mixture of bivariate Gaussian, we have two parameter spaces $\Omega=\{\Omega_{\boldsymbol{\theta}},\Omega_{\boldsymbol{\pi}}\}$ that are associated respectively to the parameters $\boldsymbol{\theta}$ and $\boldsymbol{\pi}$. They could be defined as $\Omega_{\boldsymbol{\pi}}=\{[0,1]^K|K\in\mathbb{N}, \sum_{k=1}^K\pi_k=1\}$ and $\Omega_{\boldsymbol{\theta}}=\left\{(x_{C_k},y_{C_k})\in\mathbb{R}^2,\,\, \sigma_{k,x},\sigma_{k,y}>0 \right\}$ \cite{DirichletBook1,BouguilaGibbs}.
Let us also define $\boldsymbol{\boldsymbol{\phi}}\overset{\Delta}{=} \{\phi_i=\theta_k\,\,\, \text{if}\,\,\, (x_i,y_i)\in k\text{th}\,\,\,\text{cluster}, i=[1,N]\}$ as the latent vector of variables. These variables are needed to associate each measurement point to a cluster.


\par  To summarize, the measurement point could be generated from a mixture of bivariate Gaussian conditional $P_{\boldsymbol{\theta}, \boldsymbol{\pi}}(x,y)$ defined by a set of parameters ($\boldsymbol{\theta}$ and $\boldsymbol{\pi}$) that are sampled from $\Omega_{\boldsymbol{\pi}}$ and $\Omega_{\boldsymbol{\theta}}$ according to a prior distribution $G_0$. This corresponds to a Dirichlet process, denoted by $DP(\alpha,G_0)$, where $\alpha$ is a strictly positive constant that defines the process precision \cite{DirichletBook1,BouguilaGibbs}. That is,
\begin{equation}\label{eq8}
\begin{aligned}
&\pi_1,\pi_2,\ldots\sim \mathcal{D}(\alpha)\\
&\theta_1, \theta_2, \ldots \sim G_0\\
&\phi_1,\phi_2,\ldots,\phi_N|\boldsymbol{\theta},\boldsymbol{\pi} \sim \sum_{k}\pi_k\delta_{\theta_k}\\
&(x_{U_i},y_{U_i})|\boldsymbol{\phi} \sim \mathcal{N}_{\phi_i},
\end{aligned}
\end{equation}
where $\delta_{\theta_k}$ is a Dirac measure and $\mathcal{D(\alpha)}$ is the Dirichlet distribution with parameter $\alpha$. The choice of $\alpha$ will be discussed in Sec. \ref{paramterInit}.

\subsection{Model Parameters vs. Codebook Parameters}\label{parameters}
We link in this section the model parameters (i.e., means and covariance matrices) and those of the codebook elements. Moreover, the effect of constraint on the minimum guaranteed performance is analyzed. This gives insight into the final intended values, which will help in the inference process.
\subsubsection{Codebook}\label{codebookelem}
As mentioned earlier, the beams with the intended performance can be clustered into $K$ clusters based on the set of inferred parameters $\{\boldsymbol{\theta},\boldsymbol{\pi},\Phi\}$. Here, $K$ is the length of the vector $\boldsymbol{\pi}$ (or $\boldsymbol{\theta}$). The $K$ clusters can be seen as $K$ elements of the codebook. Moreover, as the mean of each cluster is by definition the point that maximizes the average similarity with the beams in the cluster, it is then judicious to consider the means of the mixture distributions $\{(x_{C_k},y_{C_k}), k\in[1,K]\}$ as the elements of the codebook.
\subsubsection{On the Performance Criteria}\label{outagecons}
One key parameter in the performance criteria is $O_{th}$ which defines the probability of having a gain higher than $\text{Gain}_{\text{Max}}-\gamma$. To better understand the impact of this element, we consider the following example. Let us assume that we obtained a set of clusters that contain $O_{th,1}\times 100$ percent of all the measurement points (e.g., $O_{th,1}=0.95$). This suggests that rare events (measurements) with percentage $100-O_{th,1}$ are neglected. Now, considering each cluster separately and evaluate their performance. Let us assume that in each cluster the gain is higher than $Gain_{Max}-\gamma$ for at least $O_{th,2}\times 100$ percent of the measurements belonging to the cluster. These suggests that the performance criterion is satisfied if $O_{th,1}\times O_{th,2}\geq O_{th}$. An example of the possible values of these thresholds is $\{O_{th,1}=0.95, O_{th,2}=0.95, O_{th}\simeq 0.9\}$.


\par Breaking down $O_{th}$ into two elements, as explained in the previous example, will help to detect problematic clusters. For instance, for a given cluster, if the probability of having the intended gain is less than $O_{th,2}$, then one can conclude that the cluster is oversized and hence shrinking the cluster is needed (more about this is provided in Sec. \ref{ModelInf}). In fact, the smaller the cluster size, higher similarity between the measurements and the mean of the cluster which implies higher probability to meet the intended gain. For the rest of the paper, we use $O_{th,1}$ and $O_{th,2}$ to denote, respectively, the target probability of having a point measurement belonging to one of the defined clusters and the minimum required probability per cluster of achieving the intended gain. We assume that these two parameters are set by the service provider as performance criteria. They shall be chosen such that $\{O_{th,1}\times O_{th,2}\leq O_{th}\}$.

%

\subsection{The Prior}\label{prior}
The last missing piece to completely define the Dirichlet process is the prior $G_0(\boldsymbol{\theta})$. Recall that the prior is the possible distribution over the means $\{(x_{C_k},y_{C_k})|k\in\mathbb{N}\}$ and the covariance matrices $\{\text{COV}_{C_k}|k\in\mathbb{N}\}$. Since the means and the covariance matrices are independent, $G_0(\boldsymbol{\theta})$ is simply the product of their individual prior distributions. Here, there are two major challenges to be addressed. First, we must define the appropriate priors, since arbitrarily choosing the prior will considerably contaminate the final distribution. Second, during the inference process (as will be discussed in Sec. \ref{ModelInf}), we need to provide a close form expression for $p(x_{i},y_{i})$, given the prior $G_0(\boldsymbol{\theta})$, i.e.,
\begin{equation}\label{eq21}
p(x_{i},y_{i}|G_0)=\int_{\boldsymbol{\theta}\in{\Omega}}p\left(x_{i},y_{i}|\boldsymbol{\theta}\right)G_0(\boldsymbol{\theta})\text{d}\boldsymbol{\theta},
\end{equation}
for each of the measurement points $(x_i,y_i)$. Therefore, we have also interest in choosing the prior distribution such that the integral in (\ref{eq21}) is tractable.

\par The codebook elements are more likely to be in ranges of directions and widths where there is high densities of the intended dominant signals. Based on this observation, a legitimate choice of the prior distribution over the means is a mixture of Gaussians where the mixture weights are high in the ranges with high dominant signals density. Let us denote the number of mixture elements by $K_0$,  the means by $\boldsymbol{m}_0=\{m_{0,1},m_{0,2},\ldots, m_{0,K_0}\}$, and the covariance matrices by $\boldsymbol{\Lambda}_0=\{\Lambda_{0,1},\Lambda_{0,2},\ldots, \Lambda_{0,K_0}\}$. We also use $\boldsymbol{\pi}_0=\{\pi_{0,1},\pi_{0,2},\ldots, \pi_{0,K_0}\}$ to denote the mixture weights vector. That is,
\begin{equation}\label{eq22}
(x_{C_k},y_{C_k})|\boldsymbol{\pi}_0,\boldsymbol{m}_0,\boldsymbol{\Lambda}_0\sim \sum_{k=1}^{K_0}\frac{1}{\pi_{0,k}}\mathcal{N}_{m_{0,k},\Lambda_{0,k}}(\bullet).
\end{equation}
The parameters $\{\boldsymbol{\pi}_0,\boldsymbol{m}_0,\boldsymbol{\Lambda}_0\}$ are called hyper-parameters and they have to be known a priori. The parameters $K_0$ and $m_0$ may be chosen from the histogram of the observations, where $K_0$ would be the number of peaks and $\boldsymbol{m}_0$ would be the 2D positions of those peaks. As for $\boldsymbol{\Lambda}_0$, it is difficult to obtain from the histogram. As such, to account for its uncertainty, we treat its elements as random matrices.


\par To make the integral in (\ref{eq21}) tractable, we link the distributions of $\text{COV}_{C_k}$ to that of $\Lambda_0$. Next, we provide the distribution of $\text{COV}_{C_k}$ that basically defines the dimensions of the clusters. The clusters will take elliptic shapes, since they are the bases of bivariate Gaussian distributions \cite{GuaussianBook}. Since, the exact dimensions of theses ellipses cannot be priori known,  $\text{COV}_{C_k}$ can be approximated with some uncertainty by the covariance matrix that corresponds to a circular shape and proportional to $\text{COV}_{0}=I_{2\times2}$ ($I_{2\times2}$ is the $2\times 2$ identity matrix.) That is the distribution of  $\text{COV}_{C_k}$ is a Wishart distribution with parameters $\text{COV}_{0}$ and of degree two that is denoted by $\mathcal{W}_{\text{COV}_{0},2}(\bullet)$ \cite{Wishart}. Indeed, the number two comes from the fact that $\text{COV}_{B_k}$ are symmetric and can be defined via two elements which are $\sigma^2_{k,x}$ and $\sigma^2_{k,y}$. That is,
\begin{equation}\label{eq23}
\text{COV}_{C_k}\sim \mathcal{W}_{\text{COV}_{0},2}(\bullet)=\frac{\exp[-tr\left(\text{COV}_{0}^{-1}\times\bullet\right)/2]}{2^2|\text{COV}_0|\Gamma_2(1)},
\end{equation}
where $tr(\bullet)$ and $|\bullet|$ denote the trace and the determinant operators, respectively.

\par In addition to the advantage of giving a good prior for $\text{COV}_{B_k}$, the Wishart distribution is well known to be a conjugate of the Gaussian distribution, which should help in getting a closed form expression for the integral in (\ref{eq21}). Let us assume that there exists a positive constant $\varpi$ such that the elements of $\frac{1}{\varpi}\Lambda_0$ follow $\mathcal{W}(\text{COV}_{0},2)$, i.e., $\frac{1}{\varpi} \Lambda_{0,k}\sim \mathcal{W}_{\text{COV}_{0},2}$. 
In this case, the prior $G_0$ can be written as
\begin{equation}\label{eq25}
\begin{aligned}
G_0(x_{C_K},y_{C_k},\text{COV}_{C_k})
=\sum_{j=1}^{K_0}\frac{1}{\pi_{0,k}}\mathcal{N}_{m_{0,j},\frac{1}{\varpi}\text{COV}_{C_k}}(x_{C_K},y_{C_k}|\varpi\text{COV}_{C_k})\times\mathcal{W}_{\text{COV}_{0},2}(\text{COV}_{C_k}).
\end{aligned}
\end{equation}

Armed with the above results, the integral in (\ref{eq21}) is viewed as a mixture of T-distributions, which can be expressed as \cite{Tdistibution}
\begin{equation}\label{eq26}
\begin{aligned}
p(x_{i},y_{i}|G_0)&=\int_{\Omega_{\boldsymbol{\theta}}}p\left(x_{i},y_{i}|x_{C_K},y_{C_k},\text{COV}_{C_k}\right) G_0(x_{C_K},y_{C_k},\text{COV}_{C_k})\text{d}x_{C_k}\text{d}y_{C_k}\text{d}\text{COV}_{C_k}\\
&=\sum_{j=1}^{K_0}\int_{\Omega_{\boldsymbol{\theta}}}p\left(x_{i},y_{i}|x_{C_K},y_{C_k},\text{COV}_{C_k}\right) \frac{1}{\pi_{0,j}}\mathcal{N}_{m_{0,j},\frac{1}{\varpi}\text{COV}_{C_k}}(x_{C_K},y_{C_k}|\frac{1}{\varpi}\text{COV}_{C_k})\\
&\times\mathcal{W}_{\text{COV}_{0},3}(\text{COV}_{C_k})\text{d}x_{C_k}\text{d}y_{C_k}\text{d}\text{COV}_{C_k}\\
&\overset{(a)}=\sum_{j=1}^{K_0}\frac{1}{\pi_{0,j}}\mathcal{T}_{m_{1,j},t,3}(x_{i},y_{i}),
\end{aligned}
\end{equation}
where $t=\frac{3\varpi}{2(1+\varpi)}\text{COV}_0^{-1}$. The equality (a) is obtained by computing the integral as shown in \cite{Prior}.

\section{Inferring the Codebook Parameters}\label{ModelInf}
\begin{figure*}
\begin{centering}
\includegraphics[scale=0.45]{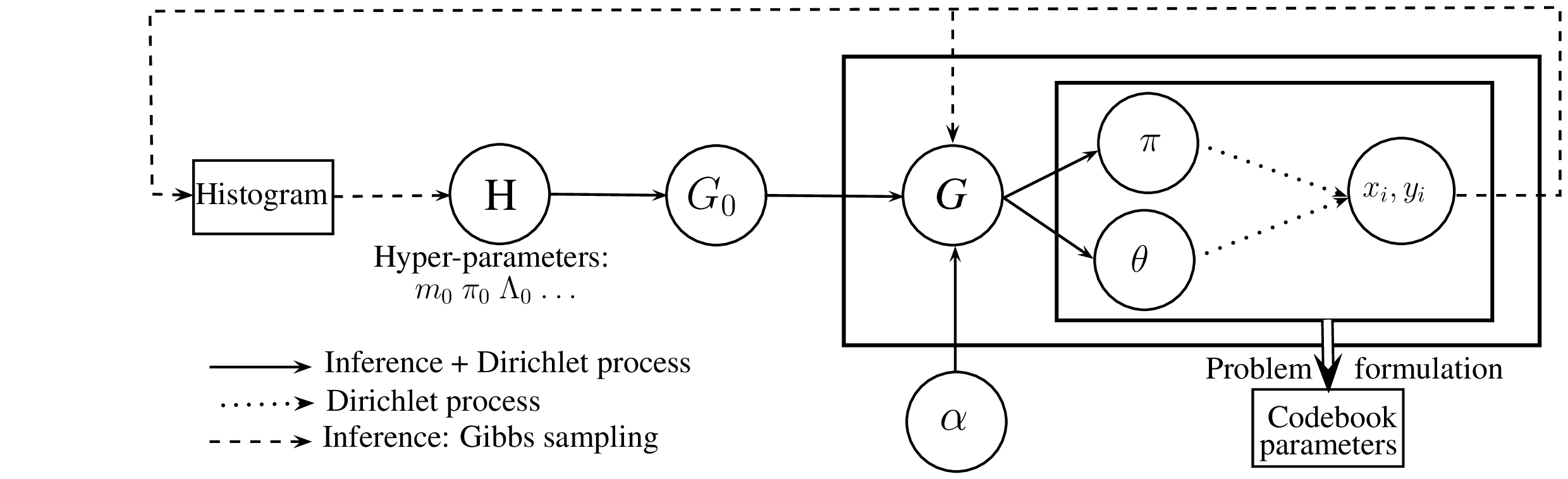}
\caption{Inference and sampling process.}
\label{diagramFig}
\end{centering}
\rule{\textwidth}{0.5pt}
\end{figure*}

The codebook parameters can be computed from the true posterior $G(\phi_n)=\sum_{k\in\mathbb{N}}\pi_k\delta_{\theta_k}(\phi_n)$. However, the true posterior is difficult to compute analytically using Bayes' rule. Nonetheless, there are inference algorithms that can provide the posterior such as the widely used Gibbs sampling approach \cite{DirichletBook1,BouguilaGibbs}. It has been shown that Gipps sampling can be executed in linear time \cite{time}. In the case of a Dirichlet process, the Gibbs sampling approach is based on the Ferguson theorem \cite{DirichletBook1,DirichletBook2,DirichletBook3}. It states that Gibbs sampling converges to the true posterior and it takes the form
\begin{equation}\label{eq28}
G\sim \frac{\alpha}{\alpha+N} G_0+ \frac{1}{\alpha+N}\sum_{n=1}^N\delta_{\phi_n}.
\end{equation}
Since only the second term in (\ref{eq28}) is considered to compute the codebook parameters, the gap to the true distribution is at most $\frac{\alpha}{\alpha+N} G_0$, which decreases as the number of samples increases. We provide a diagram in Fig. \ref{diagramFig} to summarize and emphasize the different steps of the inference and sampling processes.

%


\par We adopt in this paper, an algorithm based on the MacEachern's Gibbs sampling algorithm, which offers faster convergence compared to the naive Gibbs sampling algorithm \cite{DirichletBook1}. In the MacEachern's algorithm, two steps are executed iteratively: associating  the measurements to one of the existing clusters or generating a new one, and updating the parameters of each cluster. The MacEachern's algorithm gives results for a given process precision $\alpha$. Along with the inference process, we adjust the value of $\alpha$ using the bisection algorithm until the intended performance defined in (\ref{eq6}) is satisfied and one could not reduce the size of the codebook any more. In fact, since the number of cluster and the beamforming gain provided by the codebook increase with $\alpha$, the algorithm continue to increase $\alpha$ until the intended gain is met. Then, the size of $\alpha$ is decreased again according to bisection. The algorithm alternates between increasing and decreasing $\alpha$ until the performance criteria are met and one can not reduce $K$ further. The proposed algorithm outline is described in Algorithm \ref{AlgoG}. In the algorithm, $N_{iter}$ and $\boldsymbol{\phi}^{-i}$ denote the number of iterations and the vector that contains the elements of $\boldsymbol{\phi}$ except for the one associated to $\phi_i$.
\begin{algorithm}
\SetAlFnt{\footnotesize\sf}
\SetKwInput{KwInitialization}{Initialization}
\SetKwInput{KwStepF}{Measurements Clustering}
\SetKwInput{KwStepZ}{MacEachen Algorithm}
\SetKwInput{KwStepO}{Performance Analysis}
\SetKwInput{KwStepT}{Codebook Refining}
\SetAlFnt{\footnotesize}
\AlFnt{}

\KwInitialization{\\
~$\alpha$,~$N_{iter}$;
}
\BlankLine
\BlankLine
\KwStepF{\\
\While {Constraints on the performance $\&$ no variation on the code book size}
{
Update $\alpha$;\\
\KwStepZ{}
\For {$L=1\rightarrow N_{iter}$}
{\For {$i=1\rightarrow N$}
{
$P[\phi_i|\boldsymbol{\phi}^{-i},x_{i},y_{i}]=$
\\~$\left\{
        \begin{array}{ll}
           \frac{\alpha}{N+\alpha}\int_{\phi_{i}\in\Omega_{\boldsymbol{\theta}}}P[x_{i},y_{i}|\phi_{i}]G_0(\phi_{i})\text{d}\phi_i, & \phi_i\leftarrow\theta_{new} \\
            \frac{\sum_{j=1,j\neq i}^N\delta_{\theta_k}(\phi_j)}{N+\alpha}P[x_{i},y_{i}|\theta_k]\,\, \forall \theta_k\in\theta, & \phi_i\leftarrow \theta_k
         \end{array}
       \right.$
\\ $\phi_i\leftarrow \underset{\phi_i\in\Omega_{\boldsymbol{\theta}}}{\arg\max}P[\phi_i|\boldsymbol{\phi}^{-i},\boldsymbol{\theta},\boldsymbol{\pi},x_{i},y_{i}]$\\
 \If{$Boolean(\phi_i\leftarrow\theta_{new})== True$}
{$K\leftarrow K+1$\\
$\boldsymbol{\theta}\leftarrow\{\boldsymbol{\theta},\theta_{new}\}$
}
}
\For {$k=1\rightarrow K$}
{Update the parameter of $k$th cluster: $(x_{C_k},y_{C_k},\text{COV}_{C_k})$
}
}
\KwStepO{}
Performance analysis considering the threshold $O_{th,2}$
}
}
\KwStepT{}{
 Neglect the least probable events with sum probability of $1-O_{th,1}$.\\
Removing redundancy from the codebook.}
\caption{\small Proposed Algorithm Outline}
\label{AlgoG}
\end{algorithm}
\par Next, we discuss in detail each step in the algorithm, where we show how the results obtained in Sec. \ref{parameters} are used in the algorithm.
\subsection{Parameters Initialization}\label{paramterInit}
The proposed algorithm suggests to adjust the value of $\alpha$ using the bisection technique and hence its initial value will have only an impact on the convergence time, but not on the codebook. Nonetheless, one would anticipate a reasonably good starting value of $\alpha$ if one obtains good approximate of $K$ a priori. Indeed, considering a Dirichlet process and a number of observations $N$, the average number of generated clusters is equal to $\sum_{n=1}^{N}\frac{\alpha}{\alpha+n-1}\simeq\alpha\log\left(\frac{N}{\alpha}\right)$ \cite{DirichletBook3}. This gives
\begin{equation}\label{eq29}
\alpha=-\frac{K}{L(-K/N)},
\end{equation}
where $L(\bullet)$ is the Lambert function \cite{Lambert}. To elaborate, we consider the scenario where the transmitter and receiver are both located in the hallway of an indoor office environment. In this case, dominant signals will more likely come from the transmitter direction (LoS) and hence $K$ is expected to be around two or three. Now using $K\simeq 2\,\,\, \text{or}\,\,\, 3$, one could derive a good starting value of $\alpha$. As for the number of iterations, we set $N_{iter}$ to $50$ \cite{DirichletBook3,Bouguila1,Bouguila2,Bouguila3,Bouguila4}.

\subsection{Measurements Clustering}
During the clustering step, a measurement point is either associated to one of the existing clusters or to a new one. In fact, for each measurement point, we compute $P[\phi_i|\boldsymbol{\phi}^{-i},x_{i},y_{i}]$. A measurement point is associated to an existing cluster $C_k$ with probability
\begin{equation}\label{eq30}
\begin{aligned}
P[\phi_i=\theta_k|\boldsymbol{\phi}^{-i},x_{i},y_{i}]=\frac{\sum_{j=1,j\neq i}^N\delta_{\theta_k}(\phi_j)}{N+\alpha}P[x_{i},y_{i}|\theta_k],
\end{aligned}
\end{equation}
where $P[x_{i},y_{i}|\theta_k]\sim \mathcal{N}_{\theta_k}$, or to new cluster with probability
\begin{equation}\label{eq31}
\begin{aligned}
P[\phi_i=\theta_{new}|\boldsymbol{\phi}^{-i},x_{i},y_{i}]&=\frac{\alpha}{N+\alpha}\int_{\phi_{i}\in\Omega_{\boldsymbol{\theta}}}P[x_{i},y_{i}|\phi_{i}]G_0(\phi_{i})\text{d}\phi_i\\
&\overset{(b)}{=}\frac{\alpha}{N+\alpha}\sum_{j=1}^{K_0}\frac{1}{\pi_{0,j}}\mathcal{T}_{m_{1,j},t,3}(x_{i},y_{i}),
\end{aligned}
\end{equation}
where equality (b) comes from our derivation in (\ref{eq26}) (the T-distribution parameters are defined below (\ref{eq26}).) Then, the value of $\phi_i$ with the maximum probability will be selected. In the case when $\phi_i\leftarrow\theta_{new}$, a new randomly generated cluster will be added and the total number of clusters increases by one. The parameters of the new cluster are defined through $\theta_{new}$, which consist of mean $(x_{new},y_{new})$ and covariance matrix $\text{COV}_{new}$ that are randomly sampled from the prior.
\subsection{Clustering Parameters Update}
The means of the clusters are updated as follows.
\begin{equation}\label{eq32}
(x_{C_k},y_{C_k})=\frac{1}{\sum_{i=1}^N\delta_{\theta_k}(\phi_i)}\sum_{i=1}^N(x_{i},y_{i})\delta_{\theta_k}(\phi_i).
\end{equation}
For the covariance matrices, they are refined through multiple stages. In the first stage, the algorithm computes the most likely covariance matrix $\text{COV}_{C_k}$ given the data. That is,
\begin{subequations}\label{eq33}
  \begin{align}
    \sigma^2_{k,x}=\frac{\sum_{i=1}^N(x_{i}-x_{C_k})^2\delta_{\theta_k}(\phi_i)}{\sum_{i=1}^N\delta_{\theta_k}(\phi_i)} \\
    \sigma^2_{k,y}=\frac{\sum_{i=1}^N(y_{i}-y_{C_k})^2\delta_{\theta_k}(\phi_i)}{\sum_{i=1}^N\delta_{\theta_k}(\phi_i)}. 
  \end{align}
\end{subequations}
We then adjust the parameters of the covariance matrices to achieve the intended constraint $O_{th,2}$. It may happen that for a given cluster the probability of one of its element having a gain higher than intended one (denoted by  $\hat{O}{th,2}$) is higher or lower than $O_{th,2}$. In this case, we have interest to respectively shrink or increase the cluster size. The cluster takes an elliptical shape. Given that we have a bivariate Gaussian distribution, the surface of the ellipse that contains a given percentage of points belonging to the cluster (i..e, a given confidence interval) is proportional to $\text{pi}\sqrt{\lambda_{k,x}\lambda_{k,y}}$, where $\lambda_{k,x}$ and $\lambda_{k,y}$ are the eigenvalues of $\text{COV}_{C_k}$. To approach the intended result $O_{th,2}$ we can decrease or increase the surface covered by the cluster by the factor $\frac{\hat{O}_{th,2}}{O_{th,2}}$. The new cluster coverage becomes, $\frac{\hat{O}_{th,2}}{O_{th,2}} \text{pi} \sqrt{\lambda_{k,x}\lambda_{k,x}}=\text{pi}\sqrt{\frac{\hat{O}_{th,2}}{O_{th,2}}\lambda_{k,x}\frac{\hat{O}_{th,2}}{O_{th,2}}\lambda_{k,x}}$. This is the surface of the ellipse with covariance matrix $\frac{\hat{O}_{th,2}}{O_{th,2}}\text{COV}_{C_k}$.

\subsection{Codebook Refining}
The main objective of this step is to reduce the size of the codebook (i.e., reduce the training time) as much as possible while meeting the minimum performance criteria defined through the threshold $O_{th}=O_{th,1}\times O_{th,2}$. The output of the clustering step is a set of clusters that contain all the measurements points and each of them achieves the threshold $O_{th,2}\geq O_{th,1}$. One could eliminate the clusters containing the least probable measurement. It is to stress here that the sum of the mixture weighs ($\pi_k$) of the ignored clusters must be less than $1-O_{th,1}$.

\par In the constructed codebook, it may happen that a beam is the union of two or more narrower beams (in the codebook as well). In this case, one may keep only the narrower beams while maintaining the same performance. In fact, during the beam training, the user device checks all the codebook elements and then chooses the one that maximizes the received signal power. Knowing that the average over the union of elements is less than or equal than the maximum over the elements' averages (i.e., $\text{mean}(A,B,C)\leq \max \{\text{mean}(A),\text{mean}(B), \text{mean}(C)\}$), keeping only the narrower beams will maintain the same performance. Therefore, we suggest to ignore any redundant beam, i.e., a beam that is equal to the union of narrower beams.

\section{Exploiting Extra Side Information}\label{extension}
In previous sections, we assumed that only one side information is available which is the knowledge of a reference direction (e.g., geodetic direction). As the mobile devices are getting smarter, other side information could be available such as the geographic location and the distance from the BS. Such information could be exploited to cluster the users in smaller sets (e.g., located in closer vicinity) with higher similarity on their azimuthal beam design which will reduce the size of the codebook and then enhance the training time.

\par Although the proposed approach can take benefits from several side information, we elaborate the case when the user-BS distance is available. User devices could obtain such information using positioning systems such GPS and IPS \cite{LocalBookGPS,LocalNokiaIPS,LocalNokiaIPS1}. As discussed in Sec. \ref{Errordirec}, the proposed technique does not require highly accurate distance estimation, but rather a rough approximation. This stems from the fact that the proposed technique will use the distance in the logarithmic domain (called also log-distance) as it will be shown later on in this section. In this case, a $10$m error on the euclidian distance translates to an effective error of $\log(10)\simeq2.3$.

\par The main idea is to use the log-distance information in accordance with the average received power (isotropic power) in order to identify if a user is in LOS or in Non-LoS (NLOS) with the BS.  A codebook for each case scenario will be built from the measurements using similar method to the one described in detail in previous sections. Then, the appropriate codebook will be used for beam training. For each case scenario, it is intuitive that the codebook will be of size less than the one combining both scenarios and hence shorter beam training time is expected.

\par Measurements showed that a blockage could cause a loss of more than $20$dB in the received power. This suggests that for a given distance from the BS,
a user in NLoS with the BS characterises by a severe signal power drop as compared to a user having LoS with the BS. Therefore, it is possible to separate the LoS from the NLoS scenarios when the distance, the isotropic received power and the path loss model are available. Now, we have to build a mixture of two probabilistic models on the path loss using Bayesian learning: one associated to the LoS case and an other one to the NLoS case.

\par To build the model from the measurements collected a priori, we also make use of the concept of the Dirichlet process and Gibbs sampling for inference. To avoid dependency, we briefly describe the key elements to solve the problem (e.g., prior).
\begin{description}
  \item[1)]\textbf{Features:} isotropic power and the log-distance.
  \item[2)]\textbf{Adopted probabilistic model:} the path loss model by definition depicts the variation of the received power as a function of the log-distance. A widely used model is defined through a linear curve (slop and intercept) and root-mean-square deviation (RMS) that quantifies the error in the curve fitting. This is also equivalent to an univariate Gaussian distribution with mean ($\text{intercept}+\text{slop}\times \text{distance}$) and variance $\text{RMS}^2$. Based on the above discussion, the adopted model is a mixture of univariate Gaussian where the means take the form of ($\text{intercept}+\text{slop}\times \text{distance}$) and variances are defined by $\text{RMS}^2$.
  \item[3)] \textbf{Prior:} as prior, we consider Friis model that quantifies the drops in signal power as a function of the distance in a free space propagation environment \cite{DmitrymmWave}.
  \item[4)] \textbf{Inference algorithm:} we use Gibbs sampling for inference. During the phase of the cluster update, we make use of curve fitting to update the clusters' means and variances.
\end{description}

\section{Experimental Validation}\label{Results}
\subsection{Experiment Setup}
We used a narrowband sounder, transmitting a 28 GHz continuous-wave (CW) tone at $22$ dBm into a $10$ dBi horn with $55^{◦}$ half-power beamwidth in both elevation and azimuth. The receiver has a $10^◦$ (24 dBi) horn mounted on a rotating platform allowing a full angular scan every $200$ ms with $1^{◦}$ azimuthal angular sampling. The receiver records power samples at a rate of $740$ samples/sec, with a $20$ kHz receive bandwidth and effective noise ﬁgure of 5 dB. The system was calibrated to assure absolute power accuracy of 0.15 dBm. The high dynamic range of the sounder allows reliable measurements of the path loss up to 171 dB with directional antenna gains. A detailed description of the sounder can be found in \cite{DmitrymmWave}.

\par Measurements were performed in a Bell Labs building in Crawford Hill, NJ with a corridor of 110m long and 1.8m wide, and with offices and laboratories that may have different dimensions on both sides of the corridor. The obstacles such as furniture and equipment may differ considerably in number, location and size from one laboratory/office to another. The measurements were taken also during work hours where human obstacles were present randomly in the hallway, offices and laboratories. The transmitter was placed at one end of the hallway. During the measurements, the receiver was placed at different positions that include different offices/laboratories, different locations in a given office/laboratory and different places in the hallway. Measurements were also collected around the corners at the intersection of the hallways.


\par We collected measurements from around 300 different locations. For each location, we collected measurements for 10 seconds where the sounder rotates with speed 150 rounds per minute, i.e., $2.5$ round per second. As the received power is collected for each $1^{◦}$ azimuthal angular, we obtained measurements with an approximate size of $2.5\times 10 \times 360\times 300= 9000\times 300$. The receiver records power samples and their corresponding azimuthal angles. We used the geodic north as a reference direction. In practice, it is expected to have an error in estimating the reference direction. As discussed in Sec. \ref{Errordirec}, the error is moderate by being less than $10^{\circ}$ for almost all case scenarios. Therefore, while doing the measurements, we allowed a random error that can reach up to $20^{\circ}$. Having such error while meeting the intended performance, as shown in this section, demonstrates that the proposed approach works well under moderate errors in the reference direction estimation.

\subsection{Results}
%

We divide the measurements into two sets: the first one consists of $70\%$ of the total measurements and used it to build the codebook, whereas the remaining measurement points are used for validation. We use the collected data to evaluate the performance of the proposed scheme in terms of the azimuthal gain and training (i.e., search) time. We compare the resulting codebook design to that of the hierarchical beam design proposed in the literature. It is based on divide-and-conquer search process across the codebook levels \cite{mmWave10,mmWave11,mmWave12,mmWave13}. At each level, the beam that maximizes the received power and contained in the best higher-level wide beam is considered. It is intuitive that the higher is the number of levels the better is performance. However, the number of the levels strongly depends on the width of the narrowest beam that a device could perform. For instance, for a number of level equal to six, the mobile device is suppose to perform beam as small as $\frac{360}{2^{5}}\simeq 10^{\circ}$. Considering a number of level higher than six require that the device perform beam narrower than $\frac{360}{2^{6}}\simeq 5^{\circ}$ which not sounds practical. Therefore, we compare the performance of the proposed approach to hierarchical beam search technique considering a number of level equal to six. Moreover, we use the exhaustive search based technique as a benchmark. Recall that the $\text{Max}_{\text{Gain}}$ is achieved through exhaustive beam training where small step equal to $1^{\circ}$ is considered. In the following, we first consider the basic case where only a reference direction is available as side information. The case when the distance is also available is analyzed in Sec. \ref{SimuDist}.

\subsubsection{Codebook Design}\label{SimuDir}
\begin{figure*}
    \centering
    ~ 
    \begin{subfigure}[b]{0.4\textwidth}
        \includegraphics[scale=0.41]{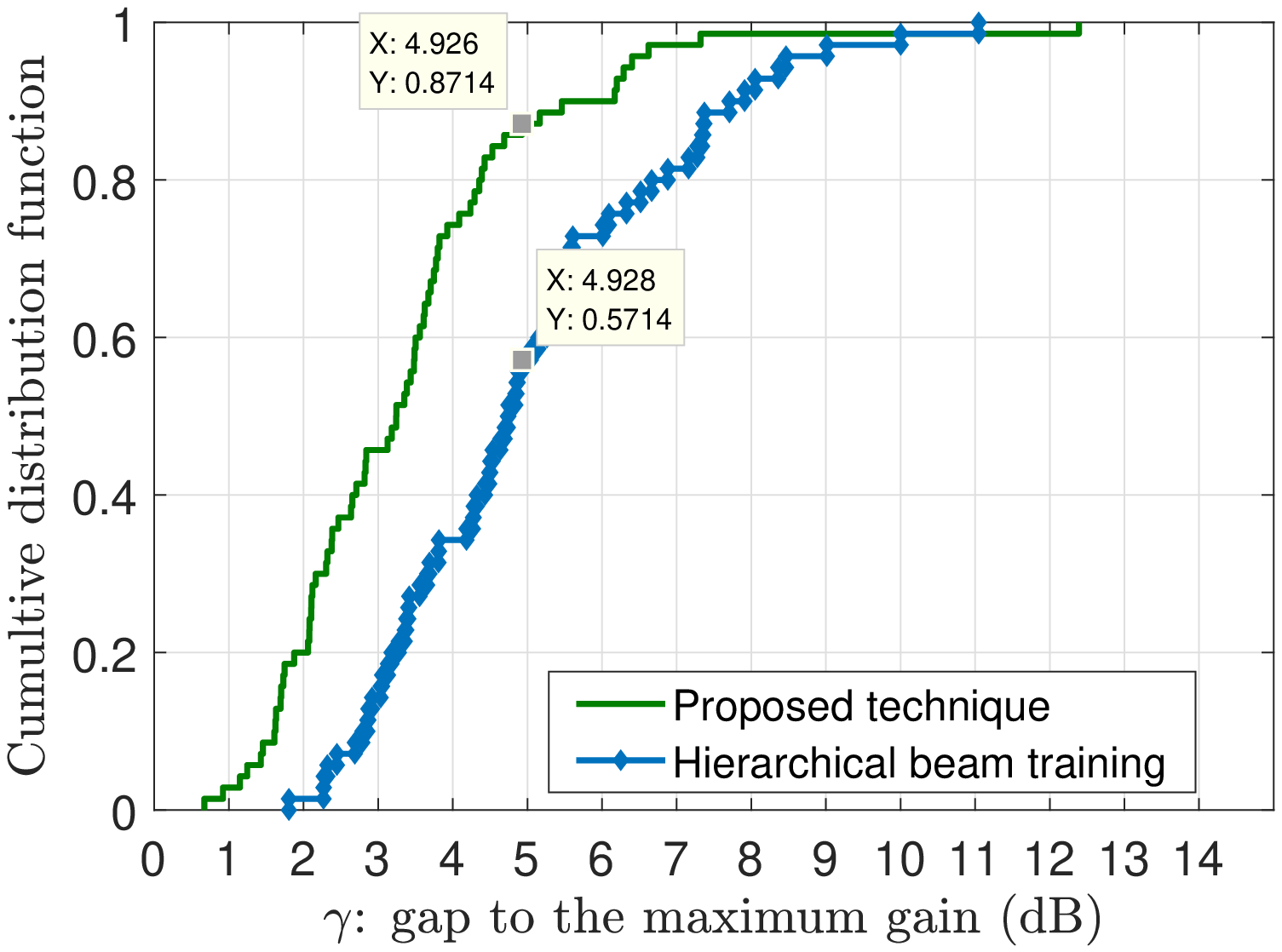}
        \caption{$\gamma=5$dB}
        \label{5dBsum}
    \end{subfigure}
\begin{subfigure}[b]{0.4\textwidth}
        \includegraphics[scale=0.41]{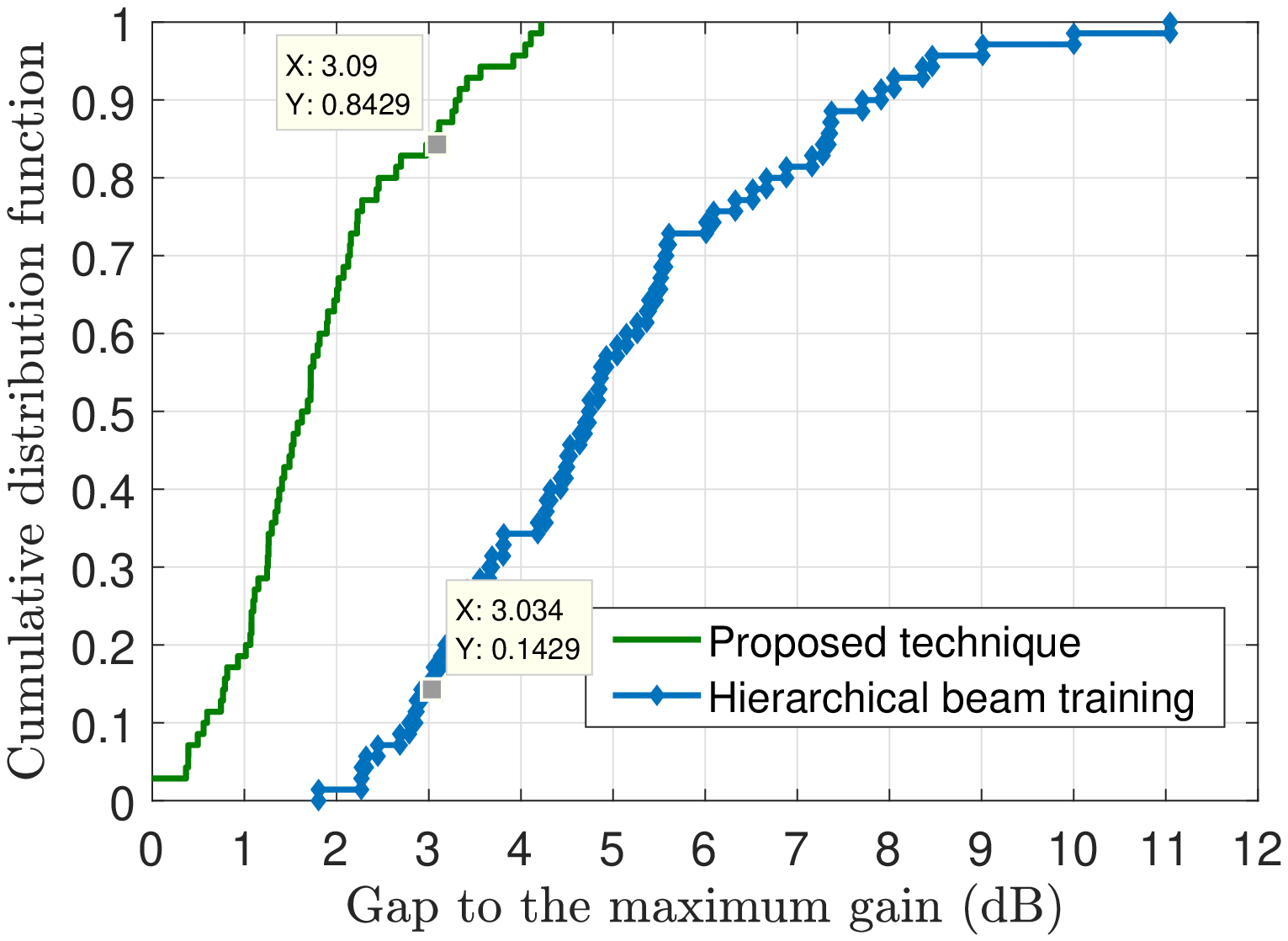}
        \caption{$\gamma=3$dB}
        \label{3dBsum}
    \end{subfigure}
    \caption{CDF of the gap to the maximum gain.}
    \rule{\textwidth}{0.5pt}
\end{figure*}

Figs. \ref{5dBsum} and \ref{3dBsum} depict the cumulative distribution function (CDF) of the gap to the maximum possible gain where $\gamma$ is chosen to be $\gamma=5$dB and $3$dB, respectively. The success rate $O_{th}$ is set to $90\%$. From the figures, it is clear that the proposed codebook almost achieves the intended performance, i.e., $Pr(\text{Gain}\geq \text{Gain}_{\text{max}}-\gamma)\geq O_{th}$. For both setups, the achievable success rate is around $85\%$ and it reaches up to $95\%$ for only one dB away from the intended gap $\gamma$. The small discrepancy to the intended rate $O_{th}=0.9$ can be explained by the fact that we are analyzing the performance of a predictor. It is therefore natural that it may not achieve the intended goal if it is facing new case scenario differs from the ones considered along with the training.

\par From the figures, we observe that hierarchical beam training approach is far away from achieving the intended goal. In fact, the success rate (i.e., achieving a gain higher than $\gamma$) is $\sim 5\%$ and $\sim 15\%$ for $\gamma=5$dB and $\gamma=3$dB, respectively. Recall that the intended goal is to provide a success rate higher than $90\%$. These clearly show the inefficiency of the hierarchical to guarantee a minimum azimuth gain.

\par In Tab. \ref{code3dB}, we provide the codebooks' elements using the proposed approach for $\gamma=5$dB and $3$dB, respectively. Recall that these codebooks are based on real measurements and hence could be used in practical systems in similar propagation scenarios. As compared to the exhaustive search approach where $360$-beams are checked, the proposed approach considerably reduces the training time by a factor $1-\frac{10}{360}\geq 95\%$.


\begin{table*}[]\scriptsize
\begin{center}
\caption{Beamforming codebook for $\gamma=5$dB and $\gamma=3$dB.}\label{code3dB}
\begin{tabular}{|c|c|c|c|c|c|c|c|c|c|c|c|c|c|c|c|c|}
\cline{1-12}
\multirow{2}{*}{$\gamma=5$dB}
  &Direction & 55 & 189 & 207 & 225 & 259 & 346 &  290 & 284 & 306 & 325 \\
  \cline{2-12}
  &Beamwidth & 22 & 23 & 21 & 25 & 27 & 20 & 19 & 24 & 19 & 25 \\
 \hline
\multirow{2}{*}{$\gamma=3$dB} &
  Direction & 349 & 264 &  279 & 186 &  199 &  242 &  327 &  211 &  230 &  248 &  292 &  304 &  57 &   70 \\
  \cline{2-16}
  &Beamwidth &  12 &   19 &   18 &  18 &   18 &   10 &   19 &   17 &   19 &   21 &   18 &   15 &   19 &   15 \\
  \hline
\end{tabular}
\end{center}
\rule{\textwidth}{0.5pt}
\end{table*}
\normalsize

\subsubsection{Codebook Design Exploiting the User-BS Distance}\label{SimuDist}\label{SimuDist}

\par
\begin{figure}[H]
\begin{centering}
\includegraphics[scale=0.45]{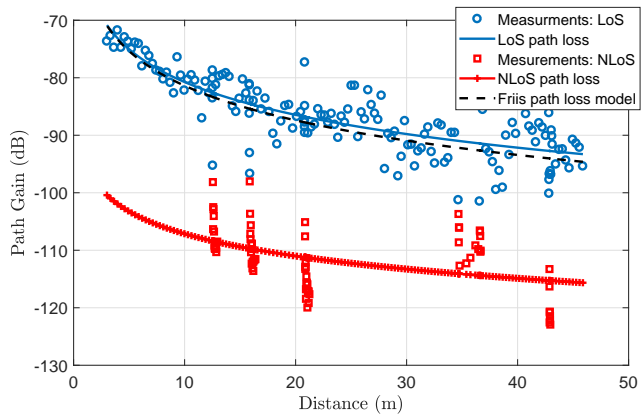}
\caption{Path loss model.}
\label{model}
\end{centering}
\end{figure}

 Fig. \ref{model} depicts the path loss models for both LoS and NLoS that are build using $70\%$ of the measurements. We used the remaining $30\%$ of measurements to validate the derived models. Using the distance and isotropic gain, the BS associates the point to one of the models in Fig. \ref{model}. We find that the inferred models provide a success rate of approximately $95\%$ while classifying a user into LoS and NLoS.

\par
\begin{figure}[H]
\begin{centering}
\includegraphics[scale=0.45]{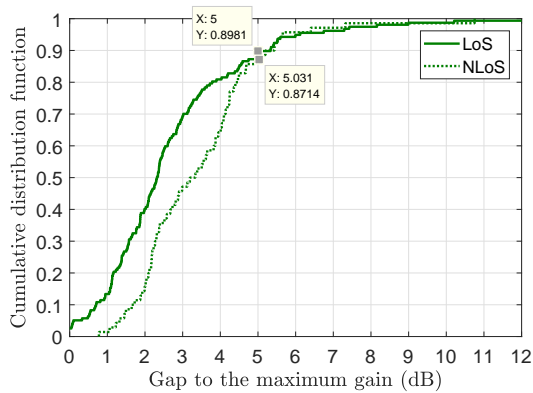}
\caption{CDF of the gap to the maximum gain for LoS and NLoS.}
\label{N+LoS}
\end{centering}
\end{figure}

 Now, after classifying a point measurement into LoS or NLoS, two codebooks are built. In Fig. \ref{N+LoS}, we provide the CDF of the gap to the maximum azimuthal gain for the NLoS and LoS scenarios. In each of scenario, we observe that each of the codebooks almost achieves the intended performance, namely, a loss lower than $5$dB for more than $90\%$ of the cases.

\begin{table*}
\begin{center}
\caption{NLoS Beamforming codebook for $\gamma=5$dB.}\label{NLoS}
\begin{tabular}{|l|c|c|c|c|c|c|c|c|c|c|c|c|c|c|}
\cline{1-5}
\multirow{2}{*}{LoS}&
  Direction & 6 & 183 &  209 \\
  \cline{2-5}
 & Beamwidth &  26 &  24 &   22  \\
  \hline
  \multirow{2}{*}{NLoS}&
  Direction &  58 &  189 &  207 &  226 &  253 &  325 &  345 &  300 & 279  \\
  \cline{2-11}
 & Beamwidth &  22 &   23 &   21 &   24 &   25 &   22 &   21 &   19 &   24 \\
  \hline
\end{tabular}
\end{center}
\rule{\textwidth}{0.5pt}
\end{table*}
\normalsize

\par In Table \ref{NLoS}, we provide the codebooks elements corresponding to LoS and NLoS, respectively. The size of the LoS and NLoS codebooks are respectively three and nine, whereas it is of size ten where both cases are combined (see Tab. \ref{code3dB}). Using the distance hence could save $\frac{10-3}{10}=70\%$ of the search time when a  user is in LoS with the BS. However, the gain is only about $10\%$ for the NLoS case. Overall, exploiting such side information could help in reducing the beam training time and the gain could be higher if more side information are available such as the location, etc.

\section{Conclusion}\label{Conclusion}
In this paper, we proposed a codebook based beamforming technique. The main feature of the proposed technique is that it does not require CSI knowledge while guaranteeing a minimum beamforming gain. It also saves more than $95\%$ of exhaustive beam training search time. The proposed technique involves using measurements that are collected from previously connected users to predict the beam designs for future connected users. In fact, the measurements are used to build a beamforming codebook that regroups the most probable beam designs containing dominant signals. We used Bayesian machine learning to cluster measurement points and to derive the appropriate codebook. The used method offers the possibility to automatically update the codebook, when changes on the physical environment occurs due, for instance, to constructions. The performance of the proposed approach is validated through a real word experiment that we conducted in indoor office environment.

\bibliographystyle{IEEEtran}
\bibliography{IEEEabrv,TWC1bib}
\end{document}